\documentclass[reprint,amsmath,amssymb,aps,pra,notitlepage,twocolumn]{revtex4-2}
\usepackage[OT4]{fontenc}
\usepackage{graphicx}
\usepackage{dcolumn}
\usepackage{bm}
\usepackage{color}

\begin{document}


\title{ Self-consistent Description of Bose-Bose Droplets:\\
Modified Gapless Hartree-Fock-Bogoliubov Method} 



\author{Pawe{\l} Zin}
\affiliation{National Centre for Nuclear Research, ul. Pasteura 7, PL-02-093 Warsaw, Poland}

\author{Maciej Pylak}
\affiliation{National Centre for Nuclear Research, ul. Pasteura 7, PL-02-093 Warsaw, Poland}

\author{Zbigniew Idziaszek}
\affiliation{Faculty of Physics, University of Warsaw, ul. Pasteura 5, PL--02--093 Warszawa, Poland}

\author{Mariusz Gajda}
\affiliation{Institute of Physics, Polish Academy of Sciences, Aleja Lotnik\'ow 32/46, PL-02-668 Warsaw, Poland}





\newcommand{\x}{{\bf r}}
\newcommand{\K}{{\bf k}}
\newcommand{\dk}{  \Delta {\bf k}}
\newcommand{\DK}{\Delta {\bf K}}
\newcommand{\KK}{{\bf K}}
\newcommand{\X}{{\bf R}}

\newcommand{\B}[1]{\mathbf{#1}} 
\newcommand{\f}[1]{\textrm{#1}} 

\newcommand{\half}{{\frac{1}{2}}}

\newcommand{\vv}{{\bf v}}
\newcommand{\p}{{\bf p}}

\newcommand{\dx}{\Delta {\bf r}}

\begin{abstract}
We define a formalism of a self-consistent description of the ground state of a weakly interacting Bose system, accounting for higher order terms in expansion of energy in the diluteness parameter. The approach is designed to be applied to a Bose-Bose mixture in a regime of weak collapse where quantum fluctuations lead to stabilization of the system and formation of quantum liquid droplets. The approach is based on the Generalized Gross -- Pitaevskii equation accounting for quantum depletion and anomalous density terms. The equation is self-consistently  coupled to  modified Bogoliubov equations. 
The modification we introduce resolves the longstanding issue of missing phonon-branch  excitations when higher order terms are included. 
Our method ensures a gapless phononic low-energy excitation spectrum, crucial to correctly account for quantum fluctuations. 
We  pay  particular attention to  the case of  droplets harmonically  confined  in some directions. The method allows to determine  the Lee-Huang-Yang-type contribution to the chemical potential of inhomogeneous droplets when the local density approximation fails.
\end{abstract}

\maketitle

\section{Introduction}

Standard theoretical description of atomic  ultracold weakly interacting Bose gas  is based on  the Gross-Pitaevskii (GP) equation \cite{Gross1961,Pitaevskii1961}  for a Bose-Einstein condensate wavefunction, and on Bogoliubov equations \cite{Bogoliubov47} giving low energy excitation spectrum. This simple approach accounts for interactions in the lowest order perturbation in a small parameter, $\sqrt{na^3} \ll 1$, where $n$ is the density of atoms and $a$ is the s-wave scattering length characterizing  interaction potential. The formalism predicts a gapless  phononic excitation spectrum in agreement  with the Hugenholtz - Pines theorem \cite{Pines}. This is a very important feature of any Bose condensate,  heralding its superfluid nature according to the Landau criterion \cite{Landau41}.  

This simple description occurs to be insufficient in some situations. Recent discoveries of quantum droplets and dipolar supersolids formed by ultracold, dilute Bosonic gases \cite{Pfau18,Petrov18,Ferrier-Barbut19,Malomed19,Bottcher_2021} proved that stability of these exotic states of matter critically depends on higher order terms in expansion of energy in the diluteness parameter.  Stabilizing role of these terms was pointed out by D.~Petrov \cite{Petrov15} predicting possibility of formation of quantum droplets.  These theoretical and experimental discoveries   triggered revival of theoretical interest in studies of ground state energy and excitation of weakly interacting Bose systems. 

These studies have a long, lasting for many decades, history \cite{Beliaev58,Girardeau59,Schick71,Popov72,Lieb63,Lieb98,Weiss04,Derezinski09,Mora09,Pilati05,Astrakharchik09}. In \cite{LHY} the two-leading terms in the expansion of diluteness parameters of the energy of hard-sphere Bosons were found. This important result showed that energy of a dilute system depends only on the s-wave scattering length $a$, not on the details of interaction potentials. Soon later, S.~T.~Beliaev \cite{Beliaev} formulated approach accounting for higher order terms in perturbation in  $\sqrt{na^3}$. This achievement paved a way towards exploration of such systems like liquid $^4$He  where interactions are not weak at all. It is commonly expected that effects of higher order terms can be significant only in strongly interacting systems.

In the pioneering paper of D.~Petrov \cite{Petrov15}, another possibility is noticed. Namely, if one  arranges such conditions that dominant mean field interaction energy vanishes,  higher order terms dominate. This might happen in a Bose-Bose mixture with repulsive intra-species interaction and  sufficiently  strong  inter-species attraction equalizing the repulsion. Mean-field description predicts a collapse of a wave-function  but  the first order correction to the interaction energy, the repulsive  Lee-Huang-Yang (LHY) energy term \cite{LHY}, might  stabilize the system at finite density.

Self-bound  quantum droplets  are formed  if interactions are in a certain range \cite{Petrov15}. The system remains dilute and weakly interacting. Its  density is of the order of $10^{15}$cm$^{-3}$. Moreover, droplets, if sufficiently large, have a constant bulk density independent on number of particles.  This property is commonly attributed  to a liquid state, although droplet's density is by orders of magnitude smaller than other liquids.  

Indeed, experiments confirmed this stabilization mechanism, \cite{Cabrera18,Tarruell18,Semeghini18,G_Ferioli20,Fort19}. Moreover, the same  mechanism occurs to be responsible for stabilizing droplets in one component gases if in addition to repulsive contact interactions, sufficiently strong long distance dipole-dipole interactions are present. Favorable conditions can be achieved in  Dysprosium or Erbium condensates, \cite{Pfau16,Pfau16a,Pfau16b,Ferlaino16}.

Calculations of LHY-like contributions to the energy, for systems in various configurations became at the center of interests of many research groups. The LHY energy is found for homogeneous Bose-Bose mixtures in 3D \cite{Petrov15}, as well as 1D or 2D geometry \cite{Petrov18} or at dimensional crossover \cite{Zin18,Ilg18}. Similar studies  give LHY energy of dipolar gases in 3D \cite{Schutzhold06,Pelster11,Pelster12} as well as in lower dimensions, at dimensional crossover \cite{Ilg18,Edler17,Jachymski18,Zin21}, or in a lattice potential \cite{Kumlin19}. 

LHY energy, once obtained, becomes a crucial ingredient of the extended GP equation as suggested in the pioneering paper of D.~Petrov \cite{Petrov15}, where  theoretical description of droplets is formulated in  two steps. In the first step, the LHY energy is found for a stable uniform system. In the second step, this  energy is added to the energy functional and extended  GP equation is obtained. The equation is applied then in a region of interactions where the mean-field description predicts instability,  manifested by  imaginary LHY energy and sound velocity of  the soft-mode. 

Evidently the approach has some drawbacks -- the LHY energy is imaginary and, in addition, corresponds to a homogeneous system. Omitting a small imaginary contribution to the LHY energy might  be excused close to the droplet formation point.   
The LHY energy term is constructed using two kinds of excitations - so called hard and soft modes. The imaginary part of the LHY energy comes from the 
soft mode contribution. 
It is very reasonable to assume that  the soft-mode excitation energy of a nonuniform stable droplet, if determined self-consistently,  is real and very small as compared to the hard mode energy. The efforts to cure imaginary excitation energies focused  on  accounting for another, in addition to  LHY term, contributions to the energy functional.  In  \cite{Hu20,Hu20b} a pairing energy was included while in \cite{Ota20}, next to the LHY energy term was added. These terms could be essential at stronger interactions, but  at the nearest proximity to the collapse, the LHY term alone is  sufficient. 

Using the LHY energy of an  uniform system  to describe a droplet has limited  applicability. This  approach is valid only if  the local density approximation is justified,  i.e. if density does not change significantly on a distance of a   healing length.  Droplets strongly squeezed in one or two directions do not meet this criterion. Their density profile changes abruptly in the confined direction. In such a case  one should use  a self-consistent approach accounting for quantum fluctuations of a  nonuniform system. In particular, a stable droplet solution of generalized GP equation should be used as a source of fluctuation field in the Bogoliubov-de Gennes  equations. Simultaneously the same  fluctuations should enter GGP equations to give a stable droplet profile. 

Here we develop the self-consistent approach capable to handle the issue --  the method that permits to  describe quantum droplets not  using  local density approximation. 
The only paper to date, considering non-homogeneous system beyond the local density approximation, assumes {\it a priori} the given density profile and cannot be generalized to  account for a self-consistent density, \cite{Ilg18}. Still there does not exist a method that enables to calculate the LHY energy using a density profile {\it modified by that LHY energy}. The need of such treatment arises in the systems with  sufficiently strong external trapping. As a  bonus, the approach should give real sound velocities.

The method defined in this paper relies on  Generalized Gross-Pitaevskii equation (GGP) describing quantum-droplet wave-function coupled to Bogoliubov equations. The GGP equation accounts  for quantum depletion and anomalous density which are to be obtained from solutions of  Bogoliubov equations self-consistently.  Unfortunately, accounting for quantum fluctuations both in GGP as well as in Bogoliubov equations meets a serious problem -- an unphysical gap in  energy spectrum  appears, so phonon-excitation branch is missing \cite{Girardeau59,Takano,Derezinski09}. Phonon excitations are crucial for low energy properties of the system.  

The issue, in a case of single component Bose gas, is solved in \cite{morgan} where Gapless Hartree-Fock-Bogoliubov approach based on the GGP equation is suggested. The approach is design   to describe frequencies and  damping of excitations  at temperatures close to  critical one.

Modified Gapless Hartree-Fock-Bogoliubov (MGHFB) method introduced here, tackles the gap problem in a different way than in \cite{morgan}, because for realistic parameters, the atoms in droplet phase  are weakly interacting  and a role of high energy contribution to quantum depletion and anomalous density is negligible. Situation is distinct from that at larger temperatures  considered by S.~A.~Morgan \cite{morgan}. We show that phononic  spectrum can be recovered provided that Bogoliubov equations are modified by introducing a relatively small (controlled by the  diluteness parameter) shift, $\delta\mu/\mu \sim \sqrt{n a^3}$, of the value of chemical potential. Instead of a value given by  mean-field wavefunction solution of the GGP equation, the eigenenergy of the zero-mode, the one which restores broken U(1) gauge symmetry, is used, \cite{Zin20zero}.  This modification  makes a colossal change -- it cures  pathological behavior of low-energy modes.  

Our self consistent approach   accounts for terms   responsible for  formation of quantum droplets --  quantum depletion and  renormalized anomalous density. The first one  describes density of noncondensed particles  in  ground state of a system, while the second measures correlations of pairs. Anomalous density is diverging with increasing  cutoff in momentum space. We introduce a physical procedure of regularization which  cancels the divergent part. The regularization procedure as well as  its numerical implementation is yet another important result of our paper. 

This paper describes  the first part of our work devoted to self-consistent description of quantum droplets.  Here we formulate  MGHFB  method -- first, for  a homogeneous case, next we focus on  inhomogeneous system, and  finally we discuss a two component inhomogeneous Bose-Bose mixture close to the transition to a droplet state.  


Results of the present paper are applied in \cite{second_paper} where we use the MGHFB method to find contribution to the chemical potential originating from quantum fluctuations for a  Bose-Bose droplet squeezed in one spatial dimension  by a harmonic potential and not confined in two remaining dimensions. This geometrical settings allows to investigate quantum fluctuations in the entire range of geometric settings: from 3D to 2D arrangements.  Local density approximation cannot be used for a tight direction in such a case.   

This  paper is organized as follows. In Sec.~\ref{sec1} we investigate a single component Bose gas to introduce main lines of our approach, first for a homogeneous system and next we generalize the method to introduce a beyond local density approach. We derive  GGP equation  introducing  quantum depletion and  renormalized anomalous density. We formulate MGHFB approach and discuss low energy excitations. 
Here we also discuss a numerical procedure to calculate  renormalized anomalous-density.
In Sec.~\ref{sec2} we use the method derived for a single component Bose gas
to specify MGHFB method for the case of a Bose-Bose mixture.  In  Appendix \ref{sec:app_A} we prove that low energy excitations are phonon-like. In Appendix \ref{semBog}, we show details of semiclassical calculations.

\section{Single component Bose gas } \label{sec1}

To systematically introduce accounting for quantum fluctuations  method of a self-consistent description of a Bose system,  we investigate  a single component case first. Calculations are less involved than in a two-component mixture case. In a derivation we use  effective  two particle interaction potential $U(\x)$. The potential serves as a tool to formulate the method. It is chosen to give a value of the  s-wave scattering length of the true interparticle potential.

For the reasons that shall become clear below, we choose the potential $U(\x)$ to be a positive bell-shape-like, extending over a distance of the order of $\sigma$, which is much larger than the s-wave scattering length, $a$, i.e. $\sigma \gg a$. The system is characterized also by two other length scales. The first one is the so called healing length given by  $\xi = \frac{\hbar}{\sqrt{m_a n g }}$ where $n$ is the density of the gas, and  $g = 4\pi \hbar^2 a/m_a$. The second one, $d$,  describes a characteristic length-scale of  density variations. The density is not uniform if the gas is confined by some external potential.  Below we  assume  that  $\sigma$ is the smallest of the two length scales, i.e. $\sigma \ll \xi,d$.

The Hamiltonian of the system:
\begin{eqnarray}
&& H = \int d \x \hat \psi^\dagger(\x) H_0 \hat \psi(\x) \nonumber 
\\
&&
+ \frac{1}{2} \int d \x  \int d \x' \, U(\x-\x')  \hat \psi^\dagger(\x) \hat \psi^\dagger(\x') \hat \psi(\x') \hat \psi(\x),
\end{eqnarray}
generates dynamics of the bosonic field $\hat \psi(\x,t)$:
\begin{eqnarray}
\label{eq1}
&& i \hbar \partial_t \hat \psi(\x,t) = H_0(\x) \hat \psi(\x,t) \nonumber
\\ 
&& + \int \mbox{d} \x' \, U(\x'-\x) \hat \psi^\dagger(\x',t) \hat \psi(\x',t) \hat \psi(\x,t),
\end{eqnarray}
where $H_0(\x)  =  -\frac{\hbar^2}{2m_a} \triangle + V(\x)$. 
In the following we consider the ground state of the system and focus on a dilute gas limit i.e. $n a^3 \ll 1$. It is known that under such conditions most of particles populate a single mode of the system -- a Bose-Einstein condensate is present. We use a standard procedure and  divide the field operator into a mean field, $\psi(\x)e^{- i \mu t } = \langle \hat \psi(\x,t) \rangle$, corresponding to a condensate, and  fluctuations, $\hat \delta(\x,t) e^{- i \mu t}$, describing out-of-condensate component: $\hat \psi(\x,t)  = \left( \psi(\x) + \hat \delta(\x,t) \right)  e^{- i \mu t}$.
In what follows we assume $\psi$ to be a real function.
Note that $ \partial_t \psi(\x) = 0$. Inserting the above into Eq.~(\ref{eq1}) we have:
\begin{widetext}
\begin{eqnarray}\label{eq22}
  i \hbar \partial_t   \hat \delta (\x,t) 
 = \left( (H_0(\x)-\mu)  
 + \int \mbox{d} \x' \,
U(\x'-\x)\left( \psi(\x') + \hat \delta^\dagger(\x',t)  \right) 
 \left( \psi(\x') + \hat \delta(\x',t)  \right)\right)
 \left( \psi(\x) + \hat \delta(\x,t)  \right), 
\end{eqnarray}
\end{widetext}
which after taking  the ground-state mean value  gives:
\begin{eqnarray}\label{GGPE}
 &&0  = \left( H_0(\x) - \mu + \tilde u (0) 
 \left( \psi^2(\x) +  2 \delta n(\x)  \right)  \right) \psi(\x)   \nonumber \\ 
&&  +
\psi(\x) \int d \x \, U(\x-\x')  m(\x',\x).
\end{eqnarray}
In the above  we introduced  anomalous density: 
\begin{equation}
 m(\x',\x) = \langle \hat \delta(\x',t) \hat \delta(\x,t) \rangle,   
\end{equation}
and quantum depletion:
\begin{equation}
 \delta n(\x',\x) = \langle \hat \delta^\dagger(\x',t)  \hat \delta(\x,t) \rangle,   
\end{equation}
In writing Eq. (\ref{GGPE}) we notice that $\langle \hat \delta \rangle =0$, and we   
neglect  the cubic term, $\sim \langle \hat \delta^\dagger\, \hat \delta \, \hat \delta \rangle$. 
Moreover, to obtain a formalism depending only on the s-wave scattering length, we introduced some further approximations:  
\begin{equation}
\int d \x' \, U(\x-\x') \psi^2(\x') \simeq  \tilde u(0)  \psi^2(\x) , 
\end{equation}
\begin{equation}
\int d \x' \, U(\x-\x') \delta n(\x',\x')  \simeq \tilde u (0) \delta n(\x) , 
\end{equation}
\begin{eqnarray}
&& \int d \x \, U(\x-\x')\psi (\x') \left( m(\x',\x) +\delta n(\x',\x) \right) \nonumber \\
&&\simeq  \tilde u(0) \delta n(\x) \psi(\x) +
\psi(\x) \int d \x \, U(\x-\x')   m(\x',\x),
\end{eqnarray}
where $\tilde u(\K)$  are Fourier components of the interaction potential, $\tilde u(\K) = \int d \x \, e^{-i \K \x} U(\x) $, in particular, $\tilde u(0) = \int d \x \, U(\x)$.
All these approximations are justified since it is assumed  that $\psi^2$ and quantum depletion $\delta n(\x)$  vary on the length scales, $\sim d$, much larger than $\sigma$. Similarly, off-diagonal quantum depletion  $\delta n(\x,\x')$   varies on distances of the order of a healing length,  $\xi \gg \sigma$. Only anomalous density $m(\x',\x)$ depends on high momenta excitation, thus changes on a smaller scale. We will discuss this issue in the following part.

In the ground state,   these quantities do not depend on time. Subtracting Eq.~(\ref{GGPE}) from Eq.~(\ref{eq22}) and preserving only terms linear in $\hat \delta$,  we get:
\begin{widetext}
\begin{equation}\label{eqn2}
i \hbar \partial_t  \hat \delta (\x,t) 
 = \left(H_0(\x)-\mu +  \tilde u (0) \psi^2(\x)  \right) \hat \delta (\x,t)  +  \psi^2(\x)  \int \mbox{d} \x' \, U(\x'-\x) \left(  \hat \delta (\x',t) +  \hat \delta^\dagger(\x',t) \right),
\end{equation}
\end{widetext}
Fluctuations $\hat \delta(\x,t)$ can be expanded into eigenmodes $u_\nu(\x)$ and $v_\nu(\x)$:
\begin{equation}\label{delta}
\hat \delta(\x,t) = \sum_\nu u_\nu(\x) e^{-i\varepsilon_\nu t/\hbar} \hat \alpha_\nu
+ v_\nu^*(\x) e^{i\varepsilon_\nu t/\hbar} \hat \alpha_\nu^\dagger,
\end{equation}
where $\hat \alpha_\nu$, $\hat \alpha_\nu^\dagger$ are annihilation and creation
operators of bosonic quasiparticles, $[\hat \alpha_\nu, \hat \alpha_{\nu'}^\dagger]=\delta_{\nu,\nu'}$.
The above bosonic commutation relation imply that:
\begin{eqnarray}\label{rel}
\sum_\nu  \left(  u_\nu(\x) u^*_\nu(\x') - v_\nu^*(\x) v_\nu(\x') \right)= \delta(\x-\x'), \\
\int \mbox{d} \x \, \left( u_\nu^*(\x) u_{\nu'}(\x) - v_\nu^*(\x) v_{\nu'}(\x) \right) = \delta_{\nu,\nu'}, \\
\int \mbox{d} \x \, \left( u_\nu^*(\x) v_{\nu'}(\x) - v_\nu^*(\x) u_{\nu'}(\x) \right) = 0.
\end{eqnarray}
Finally Bogoliubov equations are obtained:
\begin{widetext}
\begin{eqnarray}\nonumber
&& \left(H_0(\x) - \mu  + \tilde u(0) \psi^2(\x)\right) u_{\nu}(\x) 
+ \psi^2(\x) \int \mbox{d} \x' \, U(\x-\x')   u_{\nu}(\x')
+    \psi^2(\x) \int \mbox{d} \x' \, U(\x-\x')   v_{\nu}(\x') = \varepsilon_{\nu} u_{\nu}(\x)
\\ \label{Bog}
&&  \left(H_0(\x) - \mu  + \tilde u(0) \psi^2(\x)\right) v_{\nu}(\x)  + \psi^2(\x) \int \mbox{d} \x' \, U(\x-\x')   v_{\nu}(\x')
+    \psi^2(\x) \int \mbox{d} \x' \, U(\x-\x')  u_{\nu}(\x') = 
- \varepsilon_{\nu} v_{\nu}(\x).
\end{eqnarray}
\end{widetext}

Note, that  terms $m(\x',\x)$ and $\delta n(\x',\x)$ which  appear in GGP equation, Eq.~(\ref{GGPE}), are  not present in  Bogoliubov equations,  Eq.~(\ref{Bog}). This inconsistency may lead to a gap in the excitation spectrum or imaginary values of quasiparticle energies even in the ground state of the system. Both  these phenomena are nonphysical artifacts of  approximations made. We cure this problem by modification of Bogoliubov equations Eq.~(\ref{Bog}). The chemical potential $\mu$ (the one entering the GGP equation) is replaced  by another (but close) value $\mu_0$: 
\begin{equation} \label{replacement}
 \mu \to \mu_0    
\end{equation}
The analysis of the above approximation is discussed in details in 
Appendix \ref{sec:app_A}.
This replacement Eq.(\ref{replacement}) is a {\it crucial} point of our approach. It is done {\textit{ad hoc}} to compensate for the inconsistent approximations. However, this small, as it will be shown later, modification of $\mu$ in Bogoliubov equations has important consequences. In  case of  systems which are not bound (at least in one direction) it ensures a correct, gapless, phonon-like excitation spectrum (see Appendix \ref{sec:app_A} for proof of that property).

We postulate that the consistent formulation of MGHFB method should be based on the following, modified Bogoliubov equations (\ref{Bog}): 
\begin{widetext}
\begin{eqnarray} \nonumber
&& \left(H_0(\x) - \mu_0  +\tilde u(0) \psi^2(\x) \right) u_{\nu}(\x) 
+ \psi^2(\x) \int \mbox{d} \x' \, U(\x-\x')   \left( u_{\nu}(\x') +v_{\nu}(\x') \right) 
= \varepsilon_{\nu} u_{\nu}(\x)
\\ \label{BOG}
&&  \left(H_0(\x) - \mu_0  + \tilde u(0) \psi^2(\x)\right) v_{\nu}(\x)  + 
\psi^2(\x) \int \mbox{d} \x' \, U(\x-\x')   \left( u_{\nu}(\x') +v_{\nu}(\x') \right)= 
- \varepsilon_{\nu} v_{\nu}(\x).
\end{eqnarray}
\end{widetext}
where $\mu_0$  is the lowest energy solution of the following  eigenproblem:
\begin{equation}\label{GPs2}
 (H_0(\x)  +\tilde u (0) \psi^2(\x) )  u_0(\x) = \mu_0  u_0(\x),
\end{equation}
where $u_0(\x)$ and $ v_0(\x) = - u_0(\x)$ are zero energy eigenvectors recovering broken $U(1)$ gauge symmetry, \cite{Zin20zero}. 
For  completeness of the defined here approach, we remind that $\psi(\x)$ is a solution of the GGP equation, Eq.~(\ref{GGPE}):
\begin{widetext}
\begin{equation}\label{GGPEbis}
\left( H_0(\x)  + \tilde u (0)  \left( \psi^2(\x) +  2 \delta n(\x)  \right)  
+ \int d \x \, U(\x-\x')  m(\x',\x) \right) \psi(\x) = \mu \psi(\x),
\end{equation}
\end{widetext}

The equations formulated above  involve explicitly a particular form of the potential $U(\x)$, although their solutions depend on the low energy component of $U(\x)$ only. Anyway, the approach is impractical. Below we reformulate the method  starting from a homogeneous case, to pinpoint some  possible simplifications and further approximations which will grant  solutions in terms of physical quantity, namely the s-wave scattering length. These reformulation allows to generalize the formalism  to tackle inhomogeneous system. 

\subsection{Homogeneous system}

In a homogeneous case we can expand quasiparticles' eigenfunctions, $u_\nu(\x)$ and $v_\nu(\x)$, into Fourier series:
\begin{eqnarray} \label{jednorodne0}
&& u_\nu(\x)  =  \frac{1}{\sqrt{V}} \sum_\K e^{i\K\x} u_\nu(\K),\\
&& v_\nu(\x)  =  - \frac{1}{\sqrt{V}} \sum_K e^{i\K\x} v_\nu(\K),
\end{eqnarray}
so solutions of Bogoliubov equations have the form:
\begin{eqnarray}\label{unu}
&& u_\nu(\K) = \frac{1}{\sqrt{2}} \sqrt{ \frac{A_\K  }{\varepsilon_\K}  +1   },\\
\label{vnu}
&& v_\nu(\K) = \frac{1}{\sqrt{2}} \sqrt{ \frac{A_\K  }{\varepsilon_\K}  -1    }.
\end{eqnarray}
In the above $n_0 = \psi^2(\x)$ is the condensate density, $\varepsilon_\K=\sqrt{A_\K^2 - B_\K^2 }$ are quasiparticles energies, while $E_\K=\hbar^2 \K^2/2m_a$, are energies of free particles, and coefficients $A_\K$ and $B_\K$ are defined as: $A_\K = E_k + B_\K$,   $B_\K = n_0 \tilde u(\K)$. The Fourier transform of the interaction potential $\tilde u(\K) =\int  d \x \, e^{- i \K \x} U(\x)$ is assumed to be non-negative.

Inserting the above into definitions of $m$, and $\delta n$, and noticing that averaging should be performed with  quasiparticle vacuum, we arrive at the expression giving $\delta n(\x',\x)$:
\begin{eqnarray} \nonumber
&& \delta n(\x',\x)
=  \frac{1}{(2\pi)^3} \int \mbox{d} \K \, e^{i\K(\x-\x')}  v_\K^2 
\\ \label{qd}
&& =  \frac{1}{(2\pi)^3} \int  \mbox{d} \K \,  e^{i\K(\x-\x')}  \frac{1}{2} \left( \frac{E_k + n_0 u(\K)}{\varepsilon_\K}  -1 \right),
\end{eqnarray}
as well as  the pair correlation function $ m(\x',\x)$:
\begin{eqnarray}
&&  
 m(\x',\x)
= -  \frac{1}{(2\pi)^3} \int \mbox{d} \K \, e^{i\K(\x-\x')}  u_\K  v_\K
\\ \label{ano}
&& 
= -  \frac{1}{(2\pi)^3} \int \mbox{d} \K \, e^{i\K(\x-\x')} 
\frac{n_0 u(\K)}{2 \varepsilon_\K}.
\end{eqnarray}
Trying to connect the above observables with the scattering length we use Born expansion of the T-matrix which takes the form:
\begin{equation}\label{g}
 g =  {\tilde u}(0) -  \frac{1}{(2\pi)^3} \int \mbox{d} \K \, \frac{\tilde u^2(\K)}{2 E_k} + \ldots,
\end{equation}
where $g=\frac{4\pi \hbar^2}{m_a} a$ is the T-matrix and $a$ is the s-wave scattering length. 

Above, we assumed that width of potential $U(\x)$ is equal to $\sigma$. This implies that width of Fourier transform ${\tilde u}(\K)$ is  $ \sim 1/\sigma$.  This fact enables us to estimate the integral appearing above as:
\begin{equation}\label{greszta}
 \frac{1}{(2\pi)^3} \int \mbox{d} \K \, \frac{{\tilde u}^2(\K)}{2 E_k}\simeq
 \frac{  {\tilde u}^2(0)}{(2\pi)^3} \int_0^{1/\sigma}   \mbox{d} \K \, \frac{1}{2 E_k}
 =    \frac{m_a {\tilde u}^2(0) }{2 \pi^2 \hbar^2 \sigma}
\end{equation}
As a result from  Eqs.~(\ref{g}) and (\ref{greszta}) we obtain:
\begin{equation}
\left(\tilde u(0) - g \right)  \simeq   \tilde u^2(0)   \frac{m_a}{2 \pi^2 \hbar^2 \sigma}.
\end{equation}
According to our   assumption, the  range of the potential is much larger than the scattering length $a  \ll \sigma$, therefore the above can be estimated as:
\begin{eqnarray*}
\left( \tilde u(0) - g \right)   \simeq  \frac{\tilde u^2(0)}{g}  \frac{2 a}{\pi \sigma}. 
\end{eqnarray*}
The above estimation gives
\begin{equation}\label{u0_wlasnosc}
{\tilde u}(0) \simeq g
 \ \ \ \
 \left( \tilde u(0) - g \right)   \simeq  g\frac{2 a}{\pi \sigma} \ll g.
\end{equation}

The same arguments, based on the shape of interaction potentials, give that ${\tilde u}(\K) \simeq \tilde u(0)$ for  $k \ll 1/\sigma$. This implies  that the term $ \left( \frac{E_k + n_0 {\tilde u}(\K)}{\varepsilon_\K}  -1 \right)$ present in Eq.~(\ref{qd}) is proportional to $\sim 1/k^4$ for $k \gg 1/\xi$. As $\xi \gg \sigma$ thus the integral in Eq.~(\ref{qd}) converges fast enough `not to feel' the shape of the potential $\tilde u(\K)$ but only the value $\tilde u(0) \simeq g$.
As a  result  $\delta n(\x,\x')$ depends only on $g$ and we have 
\begin{equation}\label{dn}
\delta n = \delta n(\x,\x)=  \frac{8}{3 \sqrt{\pi}} (n_0 a)^{3/2}.
\end{equation}
In case of  dilute system, as  we deal with,  we have $n a^3 \ll 1$ thus $\delta n \ll n_0$. Additionally one may find that $\delta n(\x,\x')$ changes on a length scale equal to $\xi$ which again depends only on $a$ (and not on the particular choice of $\tilde u(\K)$). Eventually $\delta n(\x,\x')$ depends only on the s-wave scattering length.\\

Similarly, one can find that $\frac{n_0 \tilde u(\K)}{2 \varepsilon_\K}$, present in Eq.~(\ref{ano}), is proportional to $\tilde u(\K)/k^2$ for $k \gg 1/\xi$. As a result the anomalous average, $m(\x,\x)$, strongly depends on the shape of the potential $U$. However, the true quantity of interest is not the anomalous density but the  chemical potential. From Eq.~(\ref{GGPE}) it follows that  in  case of homogeneous system it takes the form:
\begin{equation}\label{muE}
\mu = u(0)\left( n_0 + 2 \delta n \right) 
+\int \mbox{d} \x' \, U(\x'-\x)  m(\x',\x).
\end{equation}
Substituting  $m(\x',\x)$ from Eq.~(\ref{ano}) into Eq.~(\ref{muE}) we obtain:
\begin{equation}\label{muu}
\mu \simeq  \tilde u(0)  n_0   + 2 g \delta n  
-  \frac{1}{(2\pi)^3} \int \mbox{d} \K \, \frac{n_0 \tilde u^2(\K)}{2 \varepsilon_\K}.
\end{equation}
In the derivation above we approximate $\tilde u(0) \simeq g$ in the expression $\tilde u(0) \delta n \simeq g \delta n$. This approximation cannot be used however, when the term  $n_0 \tilde u(0)$ is considered, because it involves large quantity, $n_0 \gg \delta n$. 
In  attempt to express $\tilde u(0) n_0$ by physical parameters, we wave to include higher order contribution to the scattering matrix, Eq.~(\ref{g}). As a result Eq.~(\ref{muu}) takes the form:
\begin{equation}\label{mu2}
\mu \simeq  g n_0 + 2 g \delta n + 
\frac{n_0}{(2\pi)^3} \int \mbox{d} \K \, \tilde u^2(\K) \left(  \frac{1}{2E_k} -   \frac{1}{2 \varepsilon_\K} \right).
\end{equation}
The expression entering  integral above is very similar to the term discussed in  case of quantum depletion. The integrated function of momenta $ (1/E_k - 1/\varepsilon_\K) $ converges on the scale of $1/\xi$ and therefore the integral in  Eq.(\ref{mu2}) gives a finite value depending only on $g$. This well-behaved integral is often referred to as the renormalized anomalous density $m^R$ (up to the multiplicative constant $g$): 
\begin{eqnarray} \nonumber
&& g m^R = 
\frac{n_0}{(2\pi)^3} \int \mbox{d} \K \, u^2(\K) \left(  \frac{1}{2E_k} -   \frac{1}{2 \varepsilon_\K} \right)
\\ \label{mR}
&&
\simeq \frac{g^2 n_0}{(2\pi)^3} \int \mbox{d} \K \,  \left(  \frac{1}{2E_k} -   \frac{1}{2 \varepsilon_\K} \right) =  g \frac{8}{\sqrt{\pi}} (n_0 a)^{3/2}.
\end{eqnarray}
The chemical potential, accounting for contribution  originating in  quantum fluctuations is therefore:
\begin{equation} \label{mu3}
\mu =  g ( n_0  + 2 \delta n + m^R). 
\end{equation}
Inserting into the above $\delta n$, $m^R$ from Eqs.~(\ref{dn}) and (\ref{mR})  
and using  $n = n_0 + \delta n$  we finally arrive at: 
\begin{equation}\label{mu4}
\mu = g n \left( 1 +  \frac{32}{3 \sqrt{\pi}} \sqrt{na^3} \right),
\end{equation}
where we additionally approximated $n_0 a^3 \simeq n a^3$. As we see the chemical potential depends only on the s-wave scattering length and not on any other detail of the $U(\x)$ potential. Eq.(\ref{mu4}) gives   chemical potential of the homogeneous system including  contributions from  quantum fluctuations $\Delta\mu=g n \frac{32}{3 \sqrt{\pi}} \sqrt{na^3}$. It is equal to standard expression, i.e. the Lee-Huang-Yang energy density  per atom,  $\Delta\mu=\frac{\partial e_{LHY}}{\partial n}$, where $e_{LHY}=gn^2\frac{64}{15 \sqrt{\pi}} \sqrt{na^3}$, \cite{LHY}. 
\\

It follows from the above discussion that the anomalous density can be estimated to be of the order $|m| \simeq n_0  \frac{a}{\sigma}$. As $a \ll \sigma $ we find that $|m| \ll n_0$. Moreover,  $\delta n \ll n_0$. It means that in our case the condensate density $\psi^2(\x)$ is much larger than $m(\x,\x')$ and $\delta n(\x,\x')$. 


\subsection{Inhomogeneous system}

In case of nonuniform system we are not able to give analytic expressions linking directly the renormalized anomalous density and quantum depletion to the s-wave scattering length of the interparticle potential $U(\x)$.  Our goal instead, is to formulate the approach  in a way which allows to find the quantities in question  numerically. In particular  we will give a prescription of calculating   the renormalized anomalous density  avoiding all unphysical singularities. In what follow we assume  that   crucial from the physical point of view properties of  $\delta n$ and $m$ remain valid also  in the inhomogeneous case. This means that inhomogeneity is not too strong.

The interpartical potential $U(\x)$ enters explicitly  the GGP equation, Eq.~(\ref{GGPE}), through the term  $\cal C$ :
\begin{eqnarray} 
\label{row02}
{\cal C} = \tilde  u(0)\left(|\psi(\x)|^2 + 2\delta n(\x) \right)\psi(\x)  \nonumber \\
 + \psi(\x) \int \mbox{d} \x' \,
U(\x'-\x)   m(\x',\x).
\end{eqnarray}
We proceed similarly as in the homogeneous case. We substitute $\tilde u(0)$ 
by an appropriate order of the T-matrix expansion, Eq.~(\ref{g}). We use the fist term of the series $\tilde u(0) = g$ in  terms involving  small parameter $\delta n$, while expansion up to the second order is adapted in the dominant term proportional to a condensate density:
\begin{eqnarray} \label{roww02}
&& {\cal C} \simeq  g \left( \psi^2(\x) + 2 \delta n(\x) \right)\psi(\x) 
\\ \nonumber
&& 
+ \psi(\x) 
\left(\int \mbox{d} \x' \, U(\x'-\x)   m(\x',\x)
+ \frac{\psi^2(\x)}{(2\pi)^3} \int \mbox{d} \K \, \frac{u^2(\K)}{2 E_k} \right) 
\end{eqnarray}
In the above both quantities
$  \frac{\psi^2(\x)}{(2\pi)^3} \int \mbox{d} \K \, \frac{u^2(\K)}{2 E_k}  $
and $ \int \mbox{d} \x' \,
U(\x'-\x) m(\x',\x) $ depend on $\sigma$ (i.e. on high energy modes).
In full analogy with homogeneous case we define a renormalized anomalous density:
\begin{equation}\label{rowwmR}
g m^R(\x) = \int \mbox{d} \x'  U(\x'-\x)   m(\x',\x)
+ \frac{\psi^2(\x)}{(2\pi)^3} \int \mbox{d} \K  \frac{\tilde u^2(\K)}{2 E_k}.
\end{equation}
From Eqs.~(\ref{GGPE}), (\ref{roww02}), and (\ref{rowwmR})
we obtain  GGP equation:
\begin{equation}\label{GGPEfinal}
 \big[ H_0 + g \left( \psi^2(\x) + 2 \delta n(\x)  +  m^R(\x)\right)  \big] \psi(\x) = \mu \psi(\x),
\end{equation}
which should be supplemented by   a normalization condition:
\begin{eqnarray}
\int d \x \, \left( \psi^2(\x) + \delta n(\x) \right) = N,
\end{eqnarray}
where $N$ denotes  number of atoms.

In the following part of this section we study GGP equation, Eq.~(\ref{GGPEfinal}), in depth. 
Our goal is to  formulate an approach to effectively determine  $\delta n$ and $m^R$. 
As discussed above, in homogeneous system the modes contributing to these quantities depend on low energy physics only, i.e. $\tilde u(0) \simeq g$. In Appendix \ref{semBog} we show that the same takes place in  inhomogeneous case. Here we shortly describe the main lines of the proof.

We divide the space of Bogoliubov modes into two parts: low and high energy one. We choose $E_c$ to be the energy dividing both sectors. This energy is chosen to be low enough for the low-energy modes to depend effectively only on $\tilde u(0) \simeq g$. In this energy sector the Bogoliubov equations (\ref{BOG})  can be approximated to read:
\begin{eqnarray}
&& \left(H_0  + 2 g \psi^2(\x)  \right) u_{\nu}(\x) +  g \psi^2(\x) v_\nu(\x) = (\varepsilon_{\nu}+\mu_0)
 u_{\nu}(\x),  \nonumber\\
 &&\\
&& \left(H_0  + 2 g \psi^2(\x)  \right) v_{\nu}(\x) +  g \psi^2(\x) u_\nu(\x) = - (\varepsilon_{\nu}+\mu_0) v_{\nu}(\x). \nonumber\\
\label{Bog22}
\end{eqnarray}
Similarly, the zero-mode equation, Eq.~(\ref{GPs2}), now  is:
\begin{equation}\label{GPs3}
  \left(H_0 + g \psi^2(\x)  \right) u_0(\x)= \mu_0 u_0(\x).
\end{equation}
These equations are to be solved numerically.
On the other hand, in the high energy sector we solve {\it analytically} the Bogoliubov equations (\ref{BOG})  using semiclassical approximation.
Accordingly, both quantum depletion and anomalous density are divided into low and high energy 
components, 
i.e. $n(\x)=n_L(\x)+n_H(\x)$ and  $m(\x,\x') = m_L(\x,\x') + m_H(\x,\x')$, where low energy quantum depletion is:
\begin{equation}\label{dnL}
\delta n_L(\x) = \sum_{\nu \in V_L}  |v_\nu(\x)|^2,
\end{equation}
and analogously, anomalous density: 
\begin{equation} \label{mmmL}
m_L(\x,\x') =\sum_{\nu \in V_L}  u_\nu(\x)v_\nu^*(\x').    
\end{equation}
The low energy contributions $n_L(\x)$ and $m_L(\x,\x')$ can be calculated numerically while
the high energy components can be  obtained using semicalssical approximation.

High energy part of anomalous  density, $m^R_H(\x)$, requires renormalization, silmilarly as in a homogeneous case. According to Eq.~(\ref{mRRR}) :
\begin{eqnarray}  \label{mRSEM}
&& m^R_H(\x) \simeq \frac{m_a}{2\pi^2 \hbar^2} k_c(\x)  \\ \nonumber
&& +  \psi^2(\x) \int \mbox{d} \Omega_\K  \int_{k_c(\x)}^\infty 
\frac{ k^2 \mbox{d} k }{(2\pi)^3} 
\left( \frac{1}{\frac{\hbar^2 k^2}{m_a}}  -  \frac{1}{2 \varepsilon(k,\x)  }  \right),   
\end{eqnarray}
where $k_c(\x)$ is the momentum dividing low and high energy sectors and is given by the equation $\varepsilon(k_c(\x),\x) = E_c  $, and
\begin{eqnarray}
\varepsilon(k,\x)= \sqrt{ A(\K,\x)^2 - B(\K,\x)^2}.
\end{eqnarray}
We define  $ A(\K,\x) = \left(\frac{\hbar^2 \K^2}{2m_a}  + V(\x) - \mu_0   + 2g   \psi^2(\x)\right)$ and $B(\K,\x) =  g \psi^2(\x) $. One can clearly see that momenta  $k$ giving dominant contribution to the integral in  Eq.~(\ref{mRSEM}),  are of the order of $k \sim \sqrt{m_a g \psi^2(\x)/\hbar^2}=1/\xi(\x)$. They are much smaller than $1/\sigma$ and the entire contribution depends only on  low momenta components of the interaction potential, $\tilde u(0)$. This observation has been already used to derive above formula where we explicitly wrote $g \simeq \tilde u(0)$ instead of $\tilde u(k)$ (see Eqs.~(\ref{mRRR_wczesniej}) and (\ref{mRRR})).  As a result, the renormalized anomalous density $m^R=m_L+m_H$ reads:
\begin{eqnarray}  \label{mRSEMwynik}
&& m^R(\x) \simeq  m_L(\x,\x)   + \frac{m_a}{2\pi^2 \hbar^2} k_c(\x)  + m^R_H(\x,\x).
\end{eqnarray}

Similarly the  high-momenta contribution to the quantum depletion  can be brought to the form (see Appendix \ref{semBog} for details):  
\begin{equation}\label{dnHnowe}
 \delta n_H(\x) \simeq    \int \mbox{d} \Omega_\K  \int_{k_c(\x)}^\infty 
\frac{ k^2 \mbox{d} k }{(2\pi)^3} 
\frac{1}{2} \left( \frac{A(k,\x)}{  \varepsilon(k,\x) }  -1   \right). 
\end{equation}

The above discussion not only gives a prescription how to obtain the regularized anomalous density,  $m^R$, but it is also a direct proof that a value of $m^R(\x)$  does not depend on a particular shape of the interaction potential and only on the s-wave scattering length, i.e. on $\tilde u(0) \simeq g$.  

In a view of this fact we might think about alternative approach  to calculate the regularized anomalous density $m^R$, directly from solutions of Bogoliubov equations (\ref{Bog22}) depending solely on $g$. In this approach we don't use  the decomposition into low and high energy contributions, nor use the high energy semiclassical formulae given by Eqs.~(\ref{mRSEMwynik}) and
(\ref{dnHnowe}). 
In Appendix \ref{semBog} we show that  such a direct approach  gives a closed expression
(see Eq.~(\ref{mRggg}))
\begin{eqnarray}
 &&m^R(\X) = \lim_{\dx \rightarrow 0} \bigg( m \left(\X, \dx \right) + g \psi^2(\X)  \frac{m_a}{4\pi \hbar^2 |\dx|} \bigg),
\label{mr}
\end{eqnarray}
where $m(\X,\dx) = m(\X+\frac{\dx}{2},\X-\frac{\dx}{2})$,  $\X = \frac{\x+\x'}{2}$ and $\dx = \x - \x'$, and $\lim_{\dx \rightarrow 0}\X=\x$. 
The above equation is equivalent to 
\begin{eqnarray}\label{mRdef}
m^R(\X) = \frac{\partial}{\partial  |\dx|} \left(  |\dx| m \left(\X, \dx \right)  \right)_{\dx =0}.
\end{eqnarray}
which is identical to the formula resulting from approximating  the interaction potential by the regularized Fermi-Huang zero-range pseudopotential,  $U(\x)=\delta(\x)\frac{\partial}{\partial \x} \x $, \cite{fermi,LHY}.
In this approach we simply have
\begin{equation}\label{dnLNN}
\delta n(\x) = \sum_{\nu}  |v_\nu(\x)|^2.
\end{equation}


In the above we described two possible schemes of implementing the method. The first one assumes calculation of $m^R$ from Eq.~(\ref{mr}) what in practice is restricted to situations where  solutions of Bogoliubov equations (\ref{Bog22}) are known analytically. In such case an  analytical expression of  $m^R$, might be also accessible (see \cite{Zin18} as an example).  On the other hand, if only numerical solutions of Bogoliubov equations (\ref{Bog22}) are in reach, semiclassical calculations become a solution of the problem and are  the only possibility in practice.  In  case of numerical approach only  low energy modes are available because every numerical approach uses a finite lattice spacing which introduces a high momenta cut-off.  



Some comments are in order now. The first question is if both  $m^R(\x,\x)$ and $\delta n(\x,\x)$ are finite. It follows from Eqs.~(\ref{mRSEMwynik}) and (\ref{dnHnowe}) that  high energy components of $m^R$ and $\delta n$, are finite indeed. But it is not necessarily true for low energy components, $m_L$ and $\delta n_L$. Obviously they are finite if we deal with discrete energy levels because of  finite number of states at low energy sector. But if  a system is {\it not confined} spatially in  only one direction,  contributions of low energy modes  scale like $1/k$  and infrared problem appears. In such a case both $\delta n$ and $m$ are infinite and one needs to introduce Bogoliubov method using density-phase representation. This, however, is not a subject of the present paper. If a spatial confinement is missing in more than one direction the same kind of $1/k$ scaling does not lead to the infrared catastrophe. 

Now, we comment on  obtaining the renormalized anomalous density using  semiclassical method. The error of this method is rooted in replacing a sum over discreet Bogoliubov energies by an  integral over  wavevectors $\sum_\nu \to \int \mbox{d} \K$. At the lower limit of integration, which is a sphere  of radius $k_c$, there is an uncertainty  in a `smooth' connection of low (discreet) and high (continuous) momenta sectors.  In fact, the cutoff can $k_c$ be chosen anywhere between the two neighboring discreet energies $E_c$, and  $E_c+\Delta E_c$, where $\frac{\hbar^2}{2m_a} k_c^2 = E_c$. Evidently it leads to some uncertainty, $\Delta k_c$, in determination of the cutoff momentum:  
\begin{equation} 
\label{cutoff}
\Delta k_c  = \frac{m_a}{\hbar^2}\frac{\Delta E_c}{k_c},
\end{equation}
All this comes about to an error $\varepsilon$ in the anomalous density to be of the order: 
\begin{equation}
\label{error}
\varepsilon \sim {4 \pi  k_c^2} \Delta k_c \frac{1}{E_c} \propto \Delta k_c.     
\end{equation}
In  obtaining  Eq.~(\ref{error}) we assumed,  based on Eq.~(\ref{mRSEM}), that high-energy contribution to the anomalous density has the form, $m_H \propto  \int_{k_c} {\mbox d}\K \frac{1}{E(\K)}$. If $\Delta k_c$ decreases with  increasing the cutoff momentum $k_c$, the error goes eventually to zero. However the problem appears in the case when $\Delta k_c$ stays constant. Below we give examples of the both cases.

The first one it is a system confined in the $z$-direction by a  harmonic potential of  a frequency $\omega_z$, while it is not confined in the remaining directions.  Such system is discussed in  \cite{second_paper}. At high energies the discreet energy levels are equally spaced  $\Delta E_c = \hbar \omega_z = const.$ and  from Eq.~(\ref{cutoff}) we obtain   $\Delta k_c \propto  1/k_c$, i.e. the error goes to zero. 

In the second case we consider  a system confined by  a box-like potential of the $z$-edge size equal to $L$, with  periodic boundary conditions in this direction (considered in \cite{Zin18}). The system is not confined in the two other directions. In such a case $\Delta E_c \propto k_c$ and Eq.~(\ref{cutoff}) gives a constant cutoff-independent error. In this case it turns out that to get a correct result, the value of $k_c$ has to be chosen exactly at the middle between the two energy levels i.e. $k_c = \frac{2\pi}{L} \left( n_c + \frac{1}{2} \right)$. 

We use the above method in the case of harmonic trapping and therefore we do bot need to worry about the above discussed error.


\section{Bose-Bose mixture }  \label{sec2}
Having analyzed a single component Bose gas we now move to a case of  Bose-Bose mixtures where quantum droplet state might exist. 

The system is described by the Hamiltonian being a sum of a single particle Hamiltonian, $H_0$, and two-body interaction energy. The first term contains kinetic energy as well as  energy related to  external trapping potential of the two species. The two-body interaction is assumed to be of the form:
\begin{eqnarray}
&&    H_{int}=\frac{1}{2}\sum_{i=1,2}\int \mbox{d}\x  \mbox{d}\x' \hat \psi_i(\x) \hat \psi_i(\x')U_i(\x-\x')\hat \psi_i(\x) \hat \psi_i(\x') \nonumber \\
&&    +\int \mbox{d}\x \, \mbox{d}\x' \hat \psi_1(\x) \hat \psi_2(\x')U_{12}(\x-\x')\hat \psi_1(\x) \hat \psi_2(\x'),
\end{eqnarray}
where $U_i(\x)$ and $U_{12}(\x)$ are interaction potentials. We assume that the two components, denoted as '1' and '2',  have equal masses (to simplify the calculation). The standard mean field approach is based on the two coupled stationary Gross-Pitaevskii equations. 
This approach predicts  a transition  from a homogeneous solution to a state which is localized and eventually collapses (tends to infinite density). The transition occurs when interparticle interactions are properly tuned. In a simplest case of  uniform mixture of species with equal masses the instability occurs when $\sqrt{g_{11}g_{22}}+g_{12} \leq 0$. Just before the transition point, on its stable side, Bogoliubov equations \cite{Petrov15,Oles08} support two kinds of excitation. They are known as {\it soft} and {\it hard}  modes. A velocity of  soft modes  goes to zero when approaching the transition point and becomes imaginary just passed the transition. A homogeneous solution  becomes unstable in this region.  On the other hand the sound velocity of  hard modes is always real and much larger than the sound velocity of the soft modes. 

Thus,  we consider in the following {\it  only hard-modes contribution to the LHY energy}. This energy does not change significantly while crossing the transition point.  It is enough though, to calculate the LHY energy directly at the transition and apply this expression also to the systems at a proximity to this point.

We need to add that in the approach presented below, one could obtain Bogoliubov equations of the soft modes as well. And then, the energies resulting from this equations would be real. 
Still this would lead to more complicated calculations, which almost does not change the properties of the system. That is why, in what follows, soft modes are  not taken into account.




To further  simplify  calculations we assume that a two-particle interaction potential is the same for both atomic species, $U_{11} = U_{22} = U$. Moreover we assume that  both gases have equal number of atoms $N$ and they are both placed in the same external potential $V(\x)$. The standard description of excitation of such system is given by the Bogoliubov method where response of the system to small perturbations is analyzed \cite{stringari}:
\begin{equation} \label{hatpsi}
\hat \psi_i = \psi_i + \hat \delta_i,
\end{equation}
where $\psi_{1,2} = \langle \hat \psi_{1,2} \rangle$ are condensate wave-functions, $\psi_1=\psi_2$.
Due to the assumed symmetry of exchanging of the species $1 \leftrightarrow 2$, it is convenient to introduce 
\begin{eqnarray}
\label{mfmodes}
\hat \psi_\pm & = & \frac{1}{\sqrt{2}} (\hat \psi_1 \pm \hat \psi_2),\\
\hat \delta_\pm & = & \frac{1}{\sqrt{2}} (\hat \delta_1 \pm \hat \delta_2),
\end{eqnarray} 
Inserting Eq.~(\ref{hatpsi}) into the above we find that:
\begin{eqnarray}
\label{deltamodes}
\hat \psi_+ & = & \psi_+ + \hat \delta_+ , \\
\hat \psi_- & = & \hat \delta_-.
\end{eqnarray}



Introduction of fields $\hat \psi_\pm$  allows to decouple  $\hat \psi_-$ modes from  $\hat \psi_+$.  The only non-vanishing mean field is the mean field of the {\it soft} mode $\psi \equiv \psi_+$. 

We have argued that contribution of fluctuations of  soft modes $ \hat \delta_+ $ to the LHY energy can be neglected. The {\it hard} mode fluctuations $\hat \delta \equiv \hat \delta_-$ give the only important contribution to the LHY term.  Justification of the introduced above notions of  the 'soft' and 'hard' modes will become clear when we introduce  GGP equation  and Bogoliubov equations. Summarizing this discussion, at the proximity to the  transition it is sufficient to  consider  only two  fields:  the mean field of the soft mode $\psi(\x,t)$ and  fluctuations of the hard mode $\hat \delta(\x,t)$.

 


Bogoliubov equations giving quantum fluctuations, $\hat \delta$, can be obtained by  linearization of the Heisenberg equation for this operator
$\hat \delta(\x,t) = e^{-i \mu t} \left( \sum_\nu u_{\nu}(\x) e^{- i \varepsilon_{\nu} t} 
\hat \alpha_{\nu}+ v_{\nu}^*(\x)  e^{ i \varepsilon_{\nu} t} \hat \alpha_{\nu}^\dagger \right)$:
\begin{widetext}
\begin{eqnarray}\nonumber
&& (H_0 - \mu ) u_{\nu}(\x) +   \psi(\x) \int \mbox{d} \x' \, U(\x-\x')  \psi(\x') u_{\nu}(\x')
+    \psi(\x) \int \mbox{d} \x' \, U(\x-\x')  \psi(\x') v_{\nu}(\x') = \varepsilon_{\nu} u_{\nu}(\x)
\\ \label{Bog-}
\\ \nonumber
&& (H_0 - \mu ) v_{\nu}(\x) +   \psi(\x) \int \mbox{d} \x' \, U(\x-\x')  \psi(\x') v_{\nu}(\x')
+    \psi(\x) \int \mbox{d} \x' \, U(\x-\x')  \psi(\x') u_{\nu}(\x') = 
- \varepsilon_{\nu} v_{\nu}(\x).
\end{eqnarray}
\end{widetext}
We  assume $U_{12} = - U$, which is justified close to the transition point and substituted  $\hat \psi(\x,t) = e^{- i \mu t} \hat \psi(\x)$ and   $\psi(\x,t) = e^{- i \mu t} \psi(\x) $.
We choose $\psi(\x)$ to be a real function. The above equations indicate that  excitation energies $\varepsilon_{\nu}$ are proportional to   energy of repulsive interactions, $ \int \mbox{d} \x' \psi(\x) U(\x-\x') \psi(\x') \propto g \psi^2(\x)$. It does not vanish at the critical point. This justifies the name 'hard modes' coined for   excitation triggered by perturbation $\hat \delta$.



Now we turn our attention to the mean field $\psi$. Averaging the Heisenberg equation over the vacuum of  hard modes and retaining quadratic terms in $\hat \delta$ only, in particular neglecting all terms involving $\hat \delta_{+}$, we get: 
\begin{widetext}
\begin{eqnarray} \label{GP}
&& 0 = (H_0 - \mu) \psi(\x) + 
\int \mbox{d} \x' \,  U_s(\x-\x')
\left(  \psi^2(\x') + \delta n(\x',\x') \right) \psi(\x)
\\ \nonumber
&& 
 + \int \mbox{d} \x' \, U_d(\x-\x')
\left( \delta n(\x',\x)  \psi(\x') + m(\x',\x)  \psi(\x') \right),
\end{eqnarray}
\end{widetext}
where
$ \delta n(\x',\x) =  \langle \hat \delta^\dagger(\x',t)  \hat \delta(\x,t) \rangle $
, $m(\x',\x) = \langle \hat \delta(\x',t)  \hat \delta(\x,t) \rangle $,
$U_s = (U+U_{12})/2$ and $ U_d = (U-U_{12})/2$.
One can check that $\delta n$ and $m$ are indeed real and time independent functions.

The problem is analogical to a  single component gas, Eq.~(\ref{row02})  -- we have to show that Eq.~(\ref{GP}) does not depend on details of interaction potentials, but only on scattering lengths related to them.  As previously we argue that  both the density  $n(\x)=\psi^2(\x)$ as well as $\delta n(\x,\x')$ change on the length scale given by a healing length $\xi$ being much larger than a range of the interaction potentials $\sigma$ assumed to be the smallest length scale, $\xi \gg \sigma$. Therefore the sum of all terms, $\cal C$, in Eq.~(\ref{GP}) involving integrals and interaction potentials $U_s$ and $U_d$  can be simplified:
\begin{eqnarray} \nonumber
&& {\cal C}  \simeq \tilde u_s(0)    \psi^2(\x)  \psi(\x)  
 +  g \delta n(\x)  \psi(\x)
\\ \label{GP2}
 && 
 + \psi(\x) \int \mbox{d} \x' \, U_d(\x-\x') m(\x',\x). 
\end{eqnarray}
In the above we used $\delta n(\x) = \delta n(\x,\x)$ 
and approximated
$ \int \mbox{d} \x' \, U(\x-\x')  m(\x',\x)  \psi(\x') \simeq 
\psi(\x)\int \mbox{d} \x' \, U(\x-\x')  m(\x',\x) $. This term depends on the
range of the potential $\sigma$, thus on high energy modes.

Again, we notice that  the dominant  term in Eq.~(\ref{GP2}) is proportional to the density of atoms, $\psi(\x)^2 \gg \delta n(\x)$, thus while  substituting  $\tilde u_s(0) = (\tilde u(0) + \tilde u_{12}(0))/2$ by scattering length $g$ and $g_{12}$, we must use the second order Born expansion for both $U$ and $U_{12}$ potentials, Eq.~(\ref{g}).  These second order terms, depending on high energy modes, together with the  anomalous density, give  regularized anomalous density:
\begin{eqnarray} 
g m^R(\x)  &=& \left( \int  \frac{\mbox{d} \K}{(2\pi)^3}\, \frac{\tilde u^2(\K) + \tilde u_{12}^2(\K)}{4 E_k}  \right)  \psi^2(\x)  \nonumber \\
&+& \int \mbox{d} \x' \, U_d(\x-\x')  m(\x',\x),
\end{eqnarray}
depending only on low momenta $k \simeq 0$ part of the interaction potentials i.e. on their scattering length  only. At the proximity  to the critical point where $U_{12} \simeq - U$ the above
expression simplifies
\begin{eqnarray} \nonumber
g m^R(\x)  &=& \left( \int  \frac{\mbox{d} \K}{(2\pi)^3}\, \frac{\tilde u^2(\K)}{2 E_k}  \right)  \psi^2(\x) 
\\ \label{mRinne}
&+& \int \mbox{d} \x' \, U(\x-\x')  m(\x',\x)
\end{eqnarray}

This way we recover the problem of a single component inhomogeneous  Bose gas and entire discussion of the previous section does apply.

Our self-consistent description of  Bose-Bose mixture at the proximity to the transition to the droplet state is based on the GGP equation:   
\begin{equation} \label{finalGP}
0 = \left(H_0 - \mu + \frac{\delta g}{2}  \psi^2(\x)   +  g \delta n(\x) + g m^R(\x) \right)   \psi(\x), 
\end{equation}
and  Bogoliubov equations Eq.~(\ref{Bog-}) where we  set $U(\x)=g\delta(\x)$.
In Eq.~(\ref{finalGP}) we introduced $g_{12}  = - g +\delta g $. 

As before, we deal here with  problem of chemical potential $\mu$ appearing in Bogoliubov equations. To have a consistent gapless approach  we shall replace the chemical potential $\mu \to \mu_0$ in these equations. The chemical potential $\mu_0$ has  to be found from Bogoliubov equation determining  the  zero-mode wavefunction, $u_{0}(\x) = - v_{0}(\x)$, when excitation is set to zero, \cite{Zin20zero}, $\varepsilon_{0} =0$: 
\begin{equation}\label{Eqmu0}
(H_0 - \mu_0) u_{0}(\x) = 0,
\end{equation}

The replacement $\mu \to \mu_0$ ensures that excitation spectrum is gapless and amplitudes of Bogoliubov modes have correct limit at low energies. However, a time dependence of fluctuations is the same as it was assumed, i.e. $\hat \delta(t) = e^{-i \mu t} \left( \sum_{\nu} u_{\nu}(\x)e^{-i \varepsilon_{\nu}t} + v_{\nu}(\x)e^{i \varepsilon_{\nu}t}\right)$. As a result of the above discussion, Bogoliubov equations of the hard mode take the form:
\begin{eqnarray}  \label{Bog-2_1}
 \left(H_0-\mu_0+   g \psi^2(\x) \right)   u_{\nu}(\x) +   g \psi^2(\x)   v_{\nu}(\x) & = & \varepsilon_{\nu}   u_{\nu}(\x),  \label{Bog-2_2} \\
\left( H_0-\mu_0+   g \psi^2(\x) \right)   v_{\nu}(\x)  
+    g\psi^2(\x)  u_{\nu}(\x) & = & - \varepsilon_{\nu} v_{\nu}(\x).
\end{eqnarray}
Solutions of  Bogoliubov equations allow to find  quantum depletion  $\delta n(\x) = \sum_{\nu \neq 0} |v_{\nu}(\x)|^2 $ and renormalized anomalous density, $m^R(\x) = \frac{\partial}{ \partial |\dx|} \left( |\dx| m \left(\X  ,\dx \right)   \right)_{\dx \rightarrow 0}$, where $m(\x,\x') = \sum_{\nu \neq 0} u_\nu(\x) v_\nu^*(\x')$
and  $\X = \frac{\x+\x'}{2}$ and $\dx = \x-\x'$.
To formally finish  formulation of the method we supply it by a normalization condition:
\begin{equation}
    \int d \x \left( \psi^2(\x) + \delta n(\x) \right) = 2N.
\end{equation}

Here we need to add that the above method used in the case of the systems studied in \cite{Zin18}  (gas placed between a two infinite plates - uniform system and periodic boundary conditions are used) gives exactly the same results as obtained in \cite{Zin18}.

We now consider a system where numerical solution of the Bogoliubov equations (\ref{Bog-2_1}-\ref{Bog-2_2}) is a necessity. Then, as in the single component case, using semiclassical method, we obtain
\begin{eqnarray} \nonumber 
&& m^R(\x) = m_{L}(\x,\x)
+ g  \psi^2\left(\x \right)  \frac{m_a}{2\pi^2 \hbar^2} k_c(\X) 
\\ \label{mRnMieszanina}
&& 
+ 
g \psi^2(\x) \int \mbox{d} \Omega_\K  \int_{k_c(\X)}^\infty 
\frac{ k^2 \mbox{d} k }{(2\pi)^3}
\left( \frac{1}{\frac{\hbar^2 k^2}{m_a}}  -  \frac{1}{2 \varepsilon(k,\X)  }  \right)   
\end{eqnarray}
where
\begin{eqnarray}
\varepsilon(k,\x)= \sqrt{ A(\K,\x)^2 -B(\x)^2  },
\end{eqnarray} 
where  $k_c(\x)$ is given by equation $\varepsilon_{\nu}(k_c(\x),\x) = E_c  $ and $A(\K,\x) = \frac{\hbar^2 k^2}{2m_a}  + V(\x) - \mu_0   + g   \psi^2(\x) $, $B(\x) =  g \psi^2(\x)  $. Using the same method we find
\begin{eqnarray}
 \delta n_{H}(\x) \simeq    \int \mbox{d} \Omega_\K  \int_{k_c(\x)}^\infty 
\frac{ k^2 \mbox{d} k }{(2\pi)^3} 
\frac{1}{2} \left( \frac{A(k,\x)}{  \varepsilon(k,\x) }  -1   \right),
\end{eqnarray}
Note, that the problem has to be solved self-consistently because $\delta n$ and $m^R$ depend on $\psi$ which in turn is a solution of GGP equation, Eq.~(\ref{finalGP}), which involves  $\delta n$ and $m^R$ as essential ingredients.

\section{Conclusions and Future Outlook}
In this paper we formulate the method allowing for a self-consistent treatment of coupled Generalized Gross-Pitaevskii and Bogoliubov de-Gennes equations to obtain ground state and low-energy excitations of a Bose-Bose mixture accounting for higher order terms in expansion of system  energy in diluteness parameter. We argue, that while terms originating in quantum fluctuations are important at the level of Generalized Gross-Pitaevskii equations, they should be omitted in Bogoliubov-de Gennes equations. Consistently, a heuristic modification of the value of chemical potential   by an amount being of the same order as other neglected terms,  gives correct spectrum of excitation energies and corresponding eigenvectors of Bogoliubov-de Gennes equations. 


In more details, our approach is based on three  pillars which define  closed, self-consistent system of equations. The pillars are: i) Generalized  Gross-Pitaevskii equation allowing to find droplet's wavefunction $\psi$. This equation accounts for quantum fluctuations, i.e. quantum depletion and renormalized anomalous density given by solutions of  Bogoliubov-de Gennes equation. ii) Modified Bogoliubov-de Gennes equations  allowing to find quantum depletion and anomalous density. The Bogoliubov-de Gennes equations are coupled to the Generalized Gross-Pitaevskii equation via the mean field, $\psi$. Modification of  chemical potential entering Bogoliubov-de Gennes equations ensures  gapless phononic excitation spectrum. iii) Regularization of anomalous density allowing to remove a cut-off-dependent high energy contribution.

The MGHFB method defined in this paper is applied by us in  \cite{second_paper} to  find quantum contribution to the LHY chemical potential of Bose-Bose droplets confined in one direction by a harmonic potential in the whole range of geometries from 2D to 3D case, including the crossover region.

Our approach should be also very usefull to describe {\it small droplets}  squeezed by external potentials in some directions. In such case accounting for a back-action of  a modified by the LHY term droplet density on the LHY energy in a self-consistent way is necessary.

\begin{acknowledgments}
This research was funded by the (Polish) National 
Science Centre Grant No. 2017/25/B/ST2/01943 (M.P. and M.G.)
and National Science Centre Grant No. 2015/17/B/ST2/00592 (P.Z and Z.I.)
\end{acknowledgments}


\appendix


\section{Prove of existence of gapless phononic branch in  low-energy excitation spectrum}
\label{sec:app_A}

%

Here we discus such an geometric arrangement where the system is not bound in at least one spatial direction. Obviously in a trapped system we always deal with a discrete spectrum of excitation and the concept of gapless excitation is meaningless. To be specific, the single particle Hamiltonian is assumed to have the form:
\begin{equation}
    H_0(\x)= -\frac{\hbar^2}{2m_a}\Delta_{\perp} +H_0(\x_{\parallel}),
\end{equation}
i.e. the particles can freely move in some  directions, $\x_{\perp}$, while are trapped in the remaining directions, $\x_{\parallel}$:
\begin{equation}
    H_0(\x_{\parallel}) = -\frac{\hbar^2}{2m_a}\Delta_{\parallel} + V(\x_{\parallel}).
\end{equation}
Our notation is quite general, and accounts in fact for two situations, i) where atoms are confined in the $z$-direction, $V(\x_{\parallel}) = V(z)$ while they  move freely in  the $(x-y)$ plane, $\x_{\perp}=(x,y)$, and ii) when free motion is possible  in one direction only, $\x_{\perp}=z$, while a two-dimensional potential, $V(\x_{\parallel})=V(x,y)$ provides  a confinement in $(x-y)$-plane. 

Bogoliubov equations, Eq.(\ref{Bog}) for low momenta excitation can be written in the form:
\begin{eqnarray}\label{f+}
&& \left(H(\x)  + 2 \tilde u(0) \psi^2(\x)\right) f^+_\nu(\x)  =  \varepsilon_{\nu} f^-_{\nu}(\x),
\\ 
\label{f-}
&&  H(\x)  f^-_{\nu}(\x) =  \varepsilon_{\nu} f^+_{\nu}(\x),
\end{eqnarray}
where we introduced functions $f^\pm_\nu(\x)$:
\begin{equation}
f^\pm_\nu(\x) = u_\nu(\x) \pm v_\nu(\x),   
\end{equation}
and the Hamiltonian: 
\begin{equation}
\label{H_mu0}
H(\x) =  H_0(\x) - \mu_0  + \tilde u(0)  \psi^2(\x) = H_{\parallel}(\x) -\frac{\hbar^2}{2m_a}\Delta_{\perp},
\end{equation}
where $H_{\parallel}(\x)=H_0(\x_{\parallel}) -\mu_0 + \tilde u(0)  \psi^2(\x)$. The crucial point of the  MGHFB method is substitution $\mu \to \mu_0$ in  Eq.(\ref{Bog}) what leads to the Hamiltonian $H(\x)$, Eq.(\ref{H_mu0}). The chemical potential $\mu_0$ is the ground state energy of the Bogoliubov quasiparticles and has to be found (together with the eigenvector $u_0(\x)$) from 
Eq.~(\ref{GPs2}) which may be rewritten as
\begin{equation}
\label{u0}
H(\x) u_0(\x) = H_{\parallel}(\x) u_0(\x) = 0. 
\end{equation}
We remind that $u_0(\x)$ is the zero-vector of  Bogoliubov equations corresponding to $\varepsilon_{0}=0$.

Eqs.~(\ref{f+}), (\ref{f-}) can be brought to the form:
\begin{equation}
\label{Hkwadrat}
\left(H(\x)  + 2\tilde u(0) \psi^2(\x)\right) H(\x)  f^-_\nu(\x)  =  \varepsilon_{\nu}^2 f^-_{\nu}(\x).
\end{equation}
Evidently, in the case of a homogeneous system in 3D we get $\varepsilon_{\nu}^2 = E_k(E_k + 2\tilde u(0) \psi^2)$, i.e. the gapless phonon branch at low momenta.  By $E_k$ we denote the kinetic energy $E_k=\frac{\hbar^2 k^2}{2m_a}$.

Now, we are going to show that this is also true if a confinement is introduced in some spatial directions. We assume that Bogoliubov eigenvector, $f^\pm_{\nu,k}(\x)$, is a product of a plane wave propagating in free directions and a confined function corresponding to discrete energy states in directions of bound motion:
\begin{equation}
\label{g+}
    f^\pm_{\nu,k}(\x) = g^\pm_{\nu,k}(\x_{\parallel}) e^{i \K_\perp \x_\perp}.
\end{equation}
We are interested in the lowest part of the excitation spectrum, therefore $E_k =\frac{\hbar^2 k_{\perp}^2}{2m_a}$ is a small parameter:
\begin{equation}
\label{small}
    E_k =\frac{\hbar^2 k_{\perp}^2}{2m_a} \ll \tilde u(0) \psi(\x)^2.
\end{equation}
Eq.(\ref{g+}) and Eq.(\ref{small})  allow to split the `quadratic in energy' Hermitian operator on the left hand side of  Eq.(\ref{Hkwadrat}) into `large', $G(\x)$, and `small' $\delta G(\x)$ components:
\begin{eqnarray}
\label{GG}
 G(\x) & = & \left(H_{\parallel}(\x)  + 2\tilde u(0) \psi^2(\x)\right) H_{\parallel}(\x), \\  
\delta G(\x) & = & E_k \big(E_k + 2(H_{\parallel}(\x)+ \tilde u(0) \psi(\x)^2) \big),
\end{eqnarray}
and write Eq.(\ref{Hkwadrat}) as follows:
\begin{equation}
    (G(\x) + \delta G(\x)) g_{\nu,k}^-(\x) = \varepsilon_k^2 g_{\nu,k}^-(\x).
\end{equation}
Note that the lowest energy Bogoliubov mode is equal to $g^-_{0,0}(\x) = u_0(\x)$,   thus $G(\x) g^-_{0,0}(\x)= H(\x)u_0(\x) = 0$.
Modes of the lowest  excitation energy can be approximated by plane waves of momentum $\K_{\perp}$ on top of $g^-_{0,k}(\x) \approx u_0(\x)$ profile, i.e. $f^-_{0,0}=e^{i\K_{\perp}\x_{\perp}}u_0(\x)$.  
The first order of perturbation in $E_k$ gives the low energy excitation spectrum: 
\begin{equation}
\label{excit}
   \varepsilon_k^2 =  \langle g^-_{0,k}| \delta G(\x) | g^-_{0,k} \rangle = 2 E_k  \langle u_0| \tilde u(0) \psi^2 | u_0 \rangle. 
\end{equation}
The second order perturbation gives corrections proportional to  $\delta G^2$ which are of the order of $E_k^2$. This completes our proof that MGHFB method leads to phonon-like gapless low energy excitation spectrum.

Finally we want to show that substitution of chemical potential $\mu$ by $\mu_0$ is crucial for the above result.   If the value $\mu$ which results form the GGPE  is used in Bogoliubov equations, the zero mode  eigenvector $u_0(\x)$ and corresponding eigenenergy $\varepsilon_0$ are to be found from  the equation:
\begin{equation}
\label{uprim}
(H_0(\x) - \mu_0 +(\mu_0 -\mu)  + \tilde u(0)  \psi^2(\x)) u_0(\x) = \varepsilon_0 u_0(\x).
\end{equation}
Comparing the above  equation with Eq.(\ref{u0}) one finds the excitation energy of the zero mode, $\varepsilon_0 = \mu_0-\mu \equiv \delta \mu$.
It follows from Eq.(\ref{excit}),  that in in such a situation the excitation spectrum misses a phonon branch and has an energy  gap:  
\begin{equation}\label{gap}
\varepsilon_k^2 = \delta \mu(\delta \mu +2 \langle u_0| \tilde u(0) \psi^2 | u_0 \rangle)  + 2 E_k (\delta \mu + \langle u_0| \tilde u(0) \psi^2 | u_0 \rangle).
\end{equation}
The relative difference between the two chemical potential can be estimated as $\mu- \mu_0 \ll gn$, i.e. the modification is of the order quantities $g \delta n$ or $g m$ which are not accounted for in Bogoliubov equations. Therefore, replacement $\mu \to \mu_0$ is of the same order of accuracy as other approximations made.  Note however, that even if $\delta \mu$ is small, the energy gap, Eq.(\ref{gap}), is much larger. It is of the order of  
$\sqrt{g n \delta \mu}$, which is not a small parameter. 


\section{Semiclassical solutions of Bogoliubov equations} \label{semBog}

In this Appendix we analyze semiclassical solution of the Bogoliubov equations given by Eq.~(\ref{BOG}) and Eq.~(\ref{GPs2}). We consider  case of a general external potential  $V(\x)$.

In the main body of the paper we defined renormalized anomalous density $m^R$ through Eq.~(\ref{rowwmR}). Now we show that this quantity depends only on modes which "feel" only $\tilde u(0)$.

To this end we solve Bogoliubov equations given by Eq.~(\ref{BOG}) at energies high enough to use semiclassical approximation. We divide the energy sector into "low"  $L$ and "high" $H$ energ part with $E_c$ being the energy separating those parts. As a result the anomalous density composes of two parts $m = m_L + m^R$ and  Eq.~(\ref{rowwmR}) takes the form
\begin{eqnarray}\label{klo}
&&gm^R(\x) = \int d \x'\, U(\x-\x') (m_L(\x',\x)+m_H(\x',\x))\nonumber \\ 
&&+ \frac{\psi^2(\x)}{(2\pi)^3} \int \mbox{d} \K  \frac{\tilde u^2(\K)}{2 E_k}
\end{eqnarray}

We shall use semiclassical approximation to calculate $g m^R_H(\x) = \int d \x'\, U(\x-\x') m_H(\x',\x) $. We assume the following form of Bogoliubov modes:
\begin{eqnarray}
&&u_\nu(\x) = u_{\nu,en}(\x)  e^{i k(\x) {\bf e}_\K \x},  \\
&&v_\nu(\x) =  v_{\nu,en}(\x) e^{i k(\x) {\bf e}_\K \x}.  
\end{eqnarray}
where $u_{\nu,en}$ and $v_{\nu,en}$ are slowly varying (with respect to $k(\x) {\bf e}_\K \x$) envelopes. Inserting such ansatz into Eq.~(\ref{BOG}) we obtain
\begin{eqnarray}
&& A(k,\x) u_{\nu,en}(\x)   + B(\x)   v_{\nu,en}(\x)  = \varepsilon_{\nu} u_{\nu,en}(\x),
\\ 
&&  A(k,\x) v_{\nu,en}(\x)   + B(\x)   u_{\nu,en}(\x)  
= - \varepsilon_{\nu} v_{\nu,en}(\x),
\end{eqnarray}
where
\begin{eqnarray}
&&A(k,\x) = E_k  + V(\x) - \mu_0   +    (\tilde u(0) + \tilde u(k))\psi^2(\x), \\
&&B(k,\x) =   \tilde u(k) \psi^2(\x). 
\end{eqnarray}
where, to simplify notation we assume that
$\tilde u(\K)$ depends only on $|\K|$.
Solving the above we find  
\begin{eqnarray} \label{cale00NN}
&& \varepsilon_\nu(k,\x) = \sqrt{ A^2(k,\x) - |B (k,\x)|^2},
\\ \label{cale01NN}
&& u_{\nu,en}(\x) = \frac{C_\nu(\x)}{\sqrt{2}} \sqrt{\frac{A(k(\x),\x)}{\varepsilon_\nu(k(\x),\x)} +1 },
\\ \nonumber
\\ \label{cale1NN}
&&   v_{\nu,en}(\x) = - \frac{C_\nu(\x)}{\sqrt{2}} \sqrt{\frac{A(k(\x),\x)}{\varepsilon_\nu(k(\x),\x)} -1 },
\end{eqnarray}
and normalization condition takes the form $\int \mbox{d} \x \, |C_\nu(\x)|^2 =1$. The semiclassical solution however, is still not complete because we have to find  values of $\varepsilon_\nu$. This can be done using Bohr-Somerfeld quantization condition. Once  having  $\varepsilon_\nu$ we proceed in the following way. We find $k(\x)$ from equation
\begin{equation}\label{enuNN}
\varepsilon_\nu =  \varepsilon_\nu(k(\x),\x) = \sqrt{ A^2(k(\x),\x) - B^2 (\x)}
\end{equation}
where we used Eq.~(\ref{cale00NN}). Next step is to find $C_\nu(\x)$ to obtain $u_\nu(\x)$ and $v_{\nu}(\x)$. Referring to semiclassical method we find 
\begin{equation}\label{CnuNN}
|C_\nu(\x)|^2 =  \int \frac{ \mbox{d} {\bf p} }{(2\pi \hbar)^3}
 \delta \left(\varepsilon_\nu - \varepsilon_\nu({\bf p}/\hbar,\x) \right) \varrho_\nu 
\end{equation}
where $\varrho_\nu$ is the constant phase-space density which is found from normalization condition
\begin{equation}\label{cond1NN}
1 = \int \mbox{d} \x \, |C_\nu(\x)|^2 =  \int \frac{  \mbox{d} \x \mbox{d} {\bf p} }{(2\pi \hbar)^3}
 \delta \left(\varepsilon_\nu - \varepsilon_\nu({\bf p}/\hbar,\x) \right) \varrho_\nu. 
\end{equation} 	  
In the above $u_{\nu,en}$ and $v_{\nu,en}$ are still not fully defined as we have only modulus of these functions.
A direction of $\K(\x)$ is also not defined.
For our purposes one can assume $u_{\nu,en}(\x) , v_{\nu,en}(\x)  \geq 0$ and $\K(\x) = k(\x){\bf e}_\K$ where ${\bf e}_\K$ is the unit vector independent of $\x$ which we choose to have uniform distribution on the unit sphere.
\\

We have:
\begin{eqnarray} \nonumber
&& m_H(\x',\x) = \sum_{\nu \in V_H } u_\nu(\x') v_\nu^*(\x)
\\ \nonumber
&& 
=      \sum_{\nu \in V_H } u_{\nu,en}(\x') v_{\nu,en}(\x)
\exp ( i k(\x') {\bf e}_\K \x'  - i k(\x) {\bf e}_\K \x)
\\ \label{rownaniemHNN}
&& 
\simeq  \sum_{\nu \in V_H } u_{\nu,en}(\x) v_{\nu,en}(\x)
\exp ( i k(\x) {\bf e}_\K (\x'- \x) ),
\end{eqnarray}
where we approximated $u_{\nu,en}(\x') \simeq  u_{\nu,en}(\x)$ and  $k(\x') \simeq k(\x)$.
Using Eq.~(\ref{rownaniemHNN}) we find
\begin{eqnarray}\label{gmHR}
&&g \widetilde m^R_H(\x) = 
\int d \x'\, U(\x-\x') m_H(\x',\x)\nonumber \\
&&\simeq  \sum_{\nu \in V_H }  \tilde u(k) u_{\nu,en}(\x) v_{\nu,en}(\x)
\end{eqnarray}

Summation $\sum_\nu$ can be replaced (approximately) by integration  over density of states: $\sum_{\nu \in V_H} \simeq \int_{E_c}^\infty \mbox{d} \varepsilon_\nu \, \rho(\varepsilon_\nu)  \frac{1}{4\pi} \int \mbox{d} \Omega_\K  $:
\begin{eqnarray}\nonumber
&& g \widetilde m^R_H(\x) \simeq  -  \frac{1}{4\pi} \int \mbox{d} \Omega_\K  \int_{E_c}^\infty \mbox{d} \varepsilon_\nu \, \rho(\varepsilon_\nu)    |C_\nu(\x)|^2 
\\ \label{suma2NN}
&& \times  
 \frac{\tilde u^2(k) \psi^2(\x)}{2 \varepsilon_\nu}  
\end{eqnarray}
where we used Eq.~(\ref{cale00NN}), (\ref{cale01NN}) and
(\ref{cale1NN}).
Here $E_c$ denotes the energy separating  low and high energy modes and $\frac{1}{4\pi} \int \mbox{d} \Omega_\K$  denotes the integral over solid angle of the $\K$ vector.
The density of energy states is equal to
\begin{equation}\label{rhowNN}
\rho(\varepsilon_\nu) =  \int \frac{\mbox{d}\x \mbox{d} {\bf p} }{(2\pi \hbar)^3}
 \delta \left(\varepsilon_\nu - \varepsilon_\nu(p/\hbar,\x) \right). 
\end{equation}
From Eqs.~(\ref{cond1NN}) and (\ref{rhowNN}) we find 
\begin{equation}\label{ggg}
 \rho(\varepsilon_\nu) = 1/\varrho_\nu.
\end{equation}
Using Eq.~(\ref{ggg})  together with Eq.~(\ref{CnuNN}) we find that
\begin{equation}\label{Cnu1NN}
\rho(\varepsilon) |C_\nu(\x)|^2 =  \int \frac{ \mbox{d} {\bf p} }{(2\pi \hbar)^3}
 \delta \left(\varepsilon - \varepsilon_\nu(p/\hbar,\x) \right).  
\end{equation}
Inserting Eq.~(\ref{Cnu1NN}) into (\ref{suma2NN}) we find
\begin{eqnarray} \nonumber
&& g \widetilde m^R_H(\x) \simeq  -  \frac{1}{4\pi} \int \mbox{d} \Omega_\K  \int_{E_c}^\infty \mbox{d} \varepsilon_\nu \, \int \frac{ \mbox{d} {\bf p} }{(2\pi \hbar)^3}
\\ \label{suma3NN}
&&
\times \delta \left(\varepsilon - \varepsilon_\nu(p/\hbar,\x) \right) 
\frac{\tilde u^2(k) \psi^2(\x)}{2 \varepsilon_\nu}.
\end{eqnarray}
Introducing  $k = p /\hbar$ and integrating  we obtain
\begin{equation}\label{suma4NN}
 g \widetilde m^R_H(\x) \simeq  -   \int \mbox{d} \Omega_\K  \int_{k_c(\x)}^\infty 
\frac{ k^2 \mbox{d} k }{(2\pi)^3}
\frac{\tilde u^2(k) \psi^2(\x)}{2 \varepsilon_\nu(k,\x)  } 
\end{equation}

Inserting Eq.~(\ref{suma4NN}) into  Eq.~(\ref{klo}) we find
\begin{widetext}
\begin{eqnarray}
\label{gmR1}
&& gm^R(\x) \simeq \int d \x'\, U(\x-\x') m_L(\x',\x) 
-   \int \mbox{d} \Omega_\K  \int_{k_c(\x)}^\infty 
\frac{ k^2 \mbox{d} k }{(2\pi)^3}
\frac{\tilde u^2(k) \psi^2(\x)}{2 \varepsilon_\nu(k,\x)  }
+ \frac{\psi^2(\x)}{(2\pi)^3} \int \mbox{d} \K  \frac{\tilde u^2(k)}{2 E_k},
\end{eqnarray}
where we approximated  $\int d \x'\, U(\x-\x') m_L(\x',\x) \simeq \tilde u(0) m_L(\x,\x) \simeq g m_L(\x,\x)$. Noticing   $ \int \mbox{d} \K =  \int \mbox{d} \Omega_\K ( \int_0^{k_c(\x)} 
+  \int_{k_c(\x)}^\infty ) $,  we rewrite above Eq.~(\ref{gmR1}) as
\begin{eqnarray} \label{mRRR_wczesniej}
 gm^R(\x) \simeq g m_L(\x,\x) 
+ \frac{\psi^2(\x)}{(2\pi)^3}  \int \mbox{d} \Omega_\K  
\left( \int_{k_c(\x)}^\infty 
 k^2 \mbox{d} k \, \tilde u^2(k)
 \left( \frac{1}{2E_k} -  
 \frac{1}{2 \varepsilon_\nu(k,\x)  } \right)
+ 
 \int_0^{k_c(\x)} 
 k^2 \mbox{d} k \, \tilde u^2(\K) \frac{1}{2 E_k} \right).
\end{eqnarray}
\end{widetext}
We notice that if $k < k_c(\x)$  we have $\tilde u(k) \simeq \tilde u(0) \simeq g$. In addition we notice that in the integral $k > k_c(\x)$ the integrated function $\left( \frac{1}{2E_k} - \frac{1}{2 \varepsilon_\nu(k,\x)  } \right)$ has significant impart to the integral for $k$ of the order of $1/\xi(\x)$. As $\xi(\x) \gg \sigma$ therefore we may approximate in this integral $\tilde u(k)  \simeq g$. As a result the above reduces to
\begin{eqnarray}\label{mRRR}
&&   m^R(\x) \simeq   m_L(\x,\x) 
+ 
 g \psi^2(\x)  \frac{m}{2\pi^2 \hbar^2} k_c(\x) 
\\ \nonumber
\\ \nonumber
&& +
\frac{g \psi^2(\x)}{(2\pi)^3}  \int \mbox{d} \Omega_\K  \int_{k_c(\x)}^\infty 
 k^2 \mbox{d} k \, 
 \left( \frac{1}{2E_k} -  
 \frac{1}{2 \varepsilon_\nu(k,\x)  } \right)
\end{eqnarray}

Similar argumentation can be used to get the high-energy contribution to the quantum depletion, $\delta n_H(\x)$. We find:
\begin{equation}\label{dnH1}
\delta n_H(\x) \simeq    \int \mbox{d} \Omega_\K  \int_{k_c(\x)}^\infty 
\frac{ k^2 \mbox{d} k }{(2\pi)^3} 
\frac{1}{2} \left( \frac{A(k,\x)}{  \varepsilon_\nu(k,\x) }  -1   \right). 
\end{equation}
In the above the integrand is important only for such $k$, for which we may approximate $\tilde u(k) \simeq \tilde u(0) \simeq g$. This substitution gives:
\begin{eqnarray*}
&&A(k,\x) = E_k  + V(\x) - \mu_0   +  2 g\psi^2(\x), \\
&&B(k,\x) =   g \psi^2(\x) 
\\
&& \varepsilon_\nu(k,\x) = \sqrt{ A^2(k,\x) - B^2 (k,\x)}
\end{eqnarray*}

As written in the main body of the paper we want to calculate $m^R(\x)$ directly from solutions of  Bogoliubov equations (\ref{Bog22}) where we use $g$ instead of $\int d \x' \, U(\x-\x')$. In such case we may also use the semiclassical method to solve these equations. Repeating the steps as above we obtain
\begin{eqnarray*}
 m_H(\X,\dx) \simeq  -   \int \mbox{d} \Omega_\K  \int_{k_c(\X)}^\infty 
\frac{ k^2 \mbox{d} k }{(2\pi)^3}
\frac{g \psi^2(\X)}{2 \varepsilon_\nu(k,\X)  } e^{i k  {\bf e}_\K \dx}
\end{eqnarray*}
By noticing that $\frac{1}{2\varepsilon_v}=\frac{1}{\hbar^2 k^2/m_a} + (- \frac{1}{\hbar^2 k^2/m_a}+ \frac{1}{2\varepsilon_v}) $ and having in mind that we are interested in the limit $|\Delta \x| \to 0$,  we rewrite the above as
\begin{widetext}
\begin{eqnarray} \nonumber
&& m_H(\X,\dx) \simeq  - g \psi^2(\X) \int \mbox{d} \Omega_\K  \int_{k_c(\X)}^\infty 
\frac{ k^2 \mbox{d} k }{(2\pi)^3}
\frac{1}{\frac{\hbar^2 k^2}{m_a}} e^{ i k  {\bf e}_\K \dx}
+ 
g \psi^2(\X) \int \mbox{d} \Omega_\K  \int_{k_c(\X)}^\infty 
\frac{ k^2 \mbox{d} k }{(2\pi)^3}
\left( \frac{1}{2E_k}  -  \frac{1}{2 \varepsilon_\nu(k,\X)  }  \right)  
\\ \nonumber
\\ \label{noweMH}
&&
= - g  \psi^2\left(\X \right) 
\left(  \frac{m_a}{4\pi |\dx|} - \frac{m_a}{2\pi^2 \hbar^2} k_c(\X) \right)
+ 
g \psi^2(\X) \int \mbox{d} \Omega_\K  \int_{k_c(\X)}^\infty 
\frac{ k^2 \mbox{d} k }{(2\pi)^3}
\left( \frac{1}{2E_k}  -  \frac{1}{2 \varepsilon_\nu(k,\X)  }  \right).   
\end{eqnarray}
\end{widetext}
Comparing Eq.~(\ref{mRSEMwynik}) with Eq.~(\ref{noweMH})
and keeping in mind that $m(\X,\dx) = m_L(\X,\dx) + m_H(\X,\dx)$ we find that
\begin{equation}\label{mRggg}
m^R(\X) = \lim_{\dx \rightarrow 0}
\left( m(\X,\dx) + g  \psi^2\left(\X \right)  \frac{m}{4\pi |\dx|} \right)
\end{equation}


\bibliography{main}
\bibliographystyle{apsrev4-1}

\end{document}